\documentstyle[12pt]{article}

\def\text{\mbox}
\newcommand{\beq}{\begin{equation}}
\newcommand{\eeq}{\end{equation}}
\newcommand{\beqa}{\begin{eqnarray}}
\newcommand{\eeqa}{\end{eqnarray}}

\begin{document}
\topmargin 0pt
\oddsidemargin 1mm
\begin{titlepage}
\begin{flushright}
DSF-T-8/99 \\
NORDITA-99/16 HE\\
March 1999\\
\end{flushright}
\setcounter{page}{0}
\vspace{15mm}
\begin{center}
{\Large CONSISTENT OFF-SHELL TREE STRING AMPLITUDES~
\footnote{Work partially supported by the European Commission TMR programme
ERBFMRX-CT96-0045 (Nordita-Copenhagen
and Universit\`{a} di Napoli)}} 
\vspace{20mm}

{\large Antonella Liccardo} and 
{\large Franco Pezzella~\footnote{e-mail:
Name.Surname@na.infn.it}}\\
{\em Dipartimento di Scienze Fisiche, Universit\`{a} di Napoli 
and I.N.F.N., Sezione di Napoli,\\
Mostra d'Oltremare, pad. 19, I-80125 Napoli, Italy}\\
\vspace{2mm}
{\large Raffaele Marotta~\footnote{e-mail: marotta@nbivms.nbi.dk}}\\
{\em Nordita, Blegdamsvej 17, DK-2100 Copenhagen \O, Denmark}
\end{center}
\vspace{7mm}

\begin{abstract}
We give a construction of off-shell tree bosonic string amplitudes, based on
the operatorial formalism of the $N$-string Vertex, with three external 
massless states both for open and closed strings by requiring their being 
projective invariant. In particular our prescription leads, in the low-energy 
limit, to the three-gluon amplitude in the usual covariant gauge.   
\end{abstract}

\vspace{1cm}

\end{titlepage}
\newpage

One of the main reasons for studying off-shell string amplitudes is the
investigation of the field theory limit of string theories where 
the inverse string tension $\alpha^{\prime}\rightarrow 0$ \cite{BK} 
\cite{DMLRM}. A detailed
analysis of the relation between string theory and field theory is non-trivial
and potentially very powerful since string theory manages to organize
scattering amplitudes in a very compact form and in a considerably low
number of diagrams. In particular,
closed strings can be used to shed light on perturbative quantum gravity 
\cite{BDS}. 

Furthermore, off-shell string amplitudes are relevant in studying 
processes which involve interactions among D-branes \cite{P}. 

Off-shell extensions of such amplitudes have been studied a great deal 
until now \cite{BK} \cite{DMLRM} \cite{CF} $\div$ \cite{BZ} 
and different prescriptions have been given according to
the pursued approaches.   

In this letter we use the operatorial formalism of the $N$-string $g$-loop 
Vertex $V_{N;g}$ \cite{DPFLHS} for computing tree open and closed string amplitudes 
with three external massless states. This simple case results to be very
clarifying about the prescription to use for the analysis of off-shell
string physics.
    
In the framework of the same formalism, in two recent papers \cite{CMPP1}
\cite{CMPP2}, 
off-shell one-loop amplitudes
with an arbitrary number of external massless particles have been
computed both for open and closed strings.

The $N$-string $g$-loop Vertex $V_{N;g}$ depends on which local holomorphic
coordinate system $\omega_{i}$ is used around the puncture $P_{i}$, 
$i=1,...,N$, through which the $N$-th external state is inserted on the world-sheet.
One can
introduce a single holomorphic coordinate $z$ so that each local coordinate
$\omega_{i}$ can be expressed as
\beq
\omega_{i} = V^{-1}_{i}(z)
\eeq
or
\beq
z=V_{i}(z),
\eeq
where $V_{i}(z)$ is a conformal transformation.
Since it is conventional to take $\omega_{i}(P_{i})=0$, it follows
that $z_{i} \equiv z(P_{i})=V_{i}(0)$. The $N$ quantities $z_{i}$ correspond
to the so-called Koba-Nielsen variables.

A choice of a local coordinate system around the puncture $P_{i}$
can be regarded as a sort of a {\em gauge} choice. 
When $V_{N;g}$ is saturated with $N$ physical string states satisfying the
mass-shell conditions, the corresponding amplitude does not depend on 
the $V_{i}$'s. If these conditions are released, then the dependence of
$V_{N;g}$ on them is transferred
to the off-shell amplitude. This is analogous to what happens
in gauge theories, where on-shell amplitudes are gauge invariant, while
their off-shell counterparts are not.

From the gauge choice made for the local maps it depends 
the field theory {\em gauge-fixed} Lagrangian which generates the amplitudes
to be compared with the string theory ones in the low-energy
limit. In fact one does not know a priori what gauge-fixing is chosen by
the string when the field theory limit is extracted from its amplitudes.

The results obtained in ref. \cite{DMLRM} in the one-loop case show that
it is possible to perform choices of $V_{i}$ which, in the case of
open bosonic strings, reproduce the field theory amplitudes obtained
by using the background gauge.

Hence it arises the necessity to define the assumptions underlying
the choice of the functions $V_{i}$'s. 

The basic assumption we make in this work is that {\em off-shell amplitudes
have to be projective invariant}; indeed projective invariance, or M\"{o}bius 
invariance, is a crucial 
property if off-shell finiteness and factorization are required \cite{KV}. 
The choice of local maps around the punctures
has therefore to guarantee this invariance.

In the specific case of the amplitude for three massless states
it turns out that, in the limit $\alpha ^{\prime }\rightarrow 0$, 
it depends on the choice of $V_{i}$  only through the ratio
$V_{i}^{\prime \prime }(0)/(V_{i}^{\prime }(0))^{2}$. 
Requiring its being projective invariant in this limit allows
to select a set of functions of KOba-Nielsen variables $(z_{1},z_{2},z_{3})$ 
depending on one parameter. The value of this latter can be fixed by 
requiring that the sum over all the anticyclic permutations of the lowest 
order term in
$\alpha'$, providing the field theory tree scattering amplitudes for
photons, be identically zero. In this way we univocally 
determine the ratio $V_{i}^{\prime \prime }(0)/(V_{i}^{\prime}(0))^{2}$.

Furthermore, requiring that the complete amplitude, and not only
its low-energy limit, be projective invariant univocally fixes $V'_{i}(0)$
that turns out to coincide with the first derivative evaluated at $z=0$ of
the projective transformation which maps the points
$\infty,0,1$ respectively in $z_{i-1},z_{i},z_{i+1}$ and which corresponds to
the so-called Lovelace choice~\cite{L}. But we would like here to stress
that the Lovelace choice does not reproduce the value of 
$V_{i}^{\prime \prime }(0)/(V_{i}^{\prime}(0))^{2}$ compatible with our
requirement of projective invariance.

It is possible to show that the value of $V'_{i}(0)$ so found
is the one that makes the tree level string Green function to reduce
to a particle Green function in the field theory limit \cite{RS}. 

The above prescription yields an expression of the off-shell three-gluon 
string amplitude, at tree level, which reproduces the corresponding one in field theory
in the most natural gauge, i.e. the usual covariant gauge, and also
coincides with the one obtained by the background field method that is the
technique in which results concerning higher orders are 
obtained in literature \cite{DMLRM} \cite{BD}. 

Our starting point is the $N$-string $0$-loop vertex for $N$ massless
closed bosonic strings: 
\[
V_{N;0}^{c} = C_{0}<\Omega |\int \left[ dm^c\right] _{N}^{0} \exp \left\{ \frac{1}{2}%
\sum\limits_{i=1}^{N}\sqrt{\frac{\alpha ^{\prime }}{2}}p_{i}\cdot \left[ 
\sqrt{\frac{\alpha ^{\prime }}{2}}p_{i}+\alpha _{1}^{(i)}\partial _{z}+%
\bar{\alpha}_{1}^{(i)}\partial _{\bar{z}}\right] \ln \left|
V_{i}^{\prime }(z)\right| ^{2}|_{z=0} \right\}
\]
\[
\times \exp \left\{ \sum_{\stackrel{i,j=1}{i\neq j}}^{N} \left[ \sqrt{\frac{\alpha ^{\prime }}{2}%
}p_{i}+\alpha _{1}^{(i)}V_{i}^{\prime }(0)\partial _{z_{i}}+
\bar{\alpha }%
_{1}^{(i)}\bar{V_{i}^{\prime }}(0)\partial _{\bar{z}_{i}}\right] \cdot
\left[ \sqrt{\frac{\alpha ^{\prime }}{2}}p_{j}+\alpha
_{1}^{(j)}V_{j}^{\prime }(0)\partial _{z_{j}}+\bar{\alpha }_{1}^{(j)}%
\bar{V}^{\prime}_{j}(0)\partial _{\bar{z}_j}\right] \right.
\]
\begin{equation}
\left.
 \times \ln
|z_{i}-z_{j}| \right \}  \label{1}
\end{equation}
where $<\Omega |$ is the direct product of the vacuum states of the 
Fock
spaces relative to the oscillators $\alpha ^{(i)}$ and $\bar{\alpha }^{(i)}$, 
while the measure on the moduli space for a Riemann surface of genus $g=0$ is: 
\begin{equation}
\left[ dm^{c}\right] _{N}^{0}\equiv \frac{\prod\limits_{i=1}^{N}d^2z_{i}%
\smallskip
\left|z_{A}-z_{B}\right|^2\left|z_{A}-z_{C}\right|^2\left|z_{B}-z_{C}\right|^2
}{%
\prod\limits_{i=1}^{N}\left|V_{i}^{\prime
}(0)\right|^2d^2z_{A}d^2z_{B}d^2z_{C}}.  \label{2}
\end{equation}

Furthermore, the normalization constant $C_{0}$ is given by 
$4 \pi^{3} / \alpha^{'} k^{2}$, where $k$ is the gravitational
coupling constant.
In the following our convention will be to use a superscript only for 
closed string objects and not for the open string ones; it will be denoted
by $c$. Through the introduction of the tree level 
two-point Green function for closed strings, defined as: 
\begin{equation}
G^{c}(z_{i},z_{j})=\ln \left| \frac{z_{i}-z_{j}}{\sqrt{V_{i}^{\prime
}(0)V^{\prime }j(0)}}\right|,  \label{3}
\end{equation}
the operator (\ref{1}) can be rewritten as follows: 
\[
V_{N;\,0}^{c}= C_{0}<\Omega |\int \left[ dm^c\right] _{N}^{0}\exp \left\{ \frac{1}{2}%
\sum\limits_{i=1}^{N}\sqrt{\frac{\alpha ^{\prime }}{2}}p_{i} \cdot \left[ 
\sqrt{\frac{\alpha ^{\prime }}{2}}p_{i}+\alpha _{1}^{(i)}\partial _{z}+%
\bar{\alpha }_{1}^{(i)}\partial _{\bar{z}}\right] \ln \left|
V_{i}^{\prime }(z)\right| ^{2}|_{z=0} \right\}
\]
\begin{equation}
\times \exp \left\{ \sum_{\stackrel{i,j=1}{i \neq j}}^{N}\left[ \sqrt{\frac{\alpha ^{\prime }}{2}%
}p_{i}+\alpha _{1}^{(i)}V_{i}^{\prime }(0)\partial _{z_{i}}+\bar{\alpha }%
_{1}^{(i)}\bar{V_{i}^{\prime }}(0)\partial _{\bar{z}_{i}}\right] \cdot
\left[ \sqrt{\frac{\alpha ^{\prime }}{2}}p_{j}+\alpha
_{1}^{(j)}V_{j}^{\prime }(0)\partial _{z_{j}}+\bar{\alpha }_{1}^{(j)}%
\bar{V^{\prime }}_{j}(0)\partial _{\bar{z}j}\right] \right.
\]
\[
\times \left.
\left[ \frac{1
}{2}\ln |V_{i}^{\prime }(0)V_{j}^{'}(0)|+G^{c}(z_{i,}z_{j})\right] \right\} . 
\label{4}
\end{equation}

In order to get off-shell scattering amplitudes among $N$ massless 
external particles, we have to saturate $V_{N;\,0}^{c}$ 
on the $N$ corresponding states, defined by: 
\begin{eqnarray}
|\varepsilon,p>&= & \frac{k}{\pi} \varepsilon_{\lambda \rho } \alpha
_{-1}^{\lambda }\alpha _{-1}^{\rho }|0,p>   \mbox{in the
closed string case},  \label{5} \\
|\varepsilon,p> &= & g_{d} \sqrt{2 \alpha^{'}} \varepsilon_{\lambda }\alpha
_{-1}^{\lambda }|0,p>   \mbox {in the open
string case,\thinspace } \label{5a}
\end{eqnarray}
with $g_{d}$ being the gauge coupling constant of the target space Yang-Mills
theory, and to release the on-shell conditions, i.e.:
\begin{equation}
p^{2}=0\qquad \qquad \qquad \qquad \qquad \varepsilon \cdot p=0 . \label{6}
\end{equation}
At tree level, the $N$-string Vertex does not contain any operator that 
mixes the left and right sectors of the closed bosonic string, which
implies that a tree closed string amplitude can be simply factorized into
two tree open string amplitudes \cite{KLT}. 
Hence, in restricting ourselves to compute $V^{c}_{3}$, we consider  
one sector 
of the operator given in eq. (4) and
write it properly for $N=3$ in the case of open bosonic strings. We get: 
\[
V_{3;0}^{o} = C_{0}^{open}<\Omega | \int \left[ dm \right]_{3}^{0} \exp \left\{ 
\frac{1}{2}
\sum_{{i=1}}^{3} \sqrt{ 2 \alpha ^{\prime }} p_{i}\cdot \left[ 
\sqrt{2 \alpha ^{\prime }}p_{i}+\alpha _{1}^{(i)}\partial _{z} \right]
\ln  V_{i}^{\prime }(z) |_{z=0} \right\} 
\]  
\[
\times \exp \left\{  \sum_{\stackrel{i,j=1}{i<j}}^{3}\left[ \sqrt{
 2 \alpha ^{\prime }}p_{i}+\alpha _{1}^{(i)}V_{i}^{\prime }(0)
\partial _{z_{i}}\right] \cdot
\left[ \sqrt{ 2\alpha ^{\prime }}p_{j}+\alpha _{1}^{(j)}V_{j}^{\prime
}(0)\partial _{z_{j}}\right] \right.
\]
\begin{equation} 
\times \left.
\left[ \frac{1}{2}\ln (V_{i}^{\prime
}(0)V_{j}^{\prime} (0))+G(z_{i},z_{j})\right] \right\}  \label{8}
\end{equation}
with $C_{0}^{open} = 1/ g^{2}_{d} \alpha^{'2}$ and
where $G(z_{i},z_{j})$ is the open string two-point function at 
tree level, which is related to the corresponding closed string one by: 
\begin{equation}
G^{c}(z_{i},z_{j})=\frac{1}{2}\ln \frac{z_{i}-z_{j}}{\sqrt{V_{i}^{\prime
}(0)V^{\prime }j(0)}}+\frac{1}{2}\ln \frac{\bar{z}_{i}-\bar{z}_{j}}{\sqrt{%
\bar{V}_{i}^{\prime}(0)\bar{V}_{j}^{\prime}}(0)}\equiv
\frac{1}{2} \left[ G(z_{i,}z_{j})+G(\bar{z}_{i,}\bar{z}_{j}) \right].  
\label{8bis}
\end{equation}
We would like here to observe that in the three-string case the integration 
is absolutely fictitious.
Indeed, through the identification $z_{A}=z_{1}, z_{B}=z_{2},
z_{C}=z_{3}$, the measure $\left[ dm\right] _{3}^{0}$ merely
reduces to: 
\begin{equation}
\left[ dm\right] _{3}^{0}\equiv \frac{(z_{1}-z_{2})(z_{1}-z_{3})(z_{2}-z_{3})%
}{\prod\limits_{i=1}^{3}(V_{i}^{\prime }(0))}.  \label{9}
\end{equation}

We do not want now to perform any
explicit choice of the local holomorfic coordinate $V_{i}(z),$ our purpose 
being to
get some constraints on this function by requiring the projective invariance
of the amplitude. Indeed, by saturating the operator (\ref{8}) on three
states as the ones defined in (\ref{5a}), without specifying any choice
of the local holomorphic function, we get: 
\[
A_{0}^{3}= 4g_{d} \left( 2\alpha ^{\prime } \right)^{-\frac{1}{2}}
(z_{1}-z_{2})^{1+ 2\alpha ^{\prime }p_{1\cdot
}p_{2}}(z_{1}-z_{3})^{1+ 2\alpha ^{\prime }p_{1\cdot
}p_{3}}(z_{2}-z_{3})^{1+ 2\alpha ^{\prime }p_{2\cdot }p_{3}} 
e^{\sum \limits_{i=1}^{3} \alpha ^{\prime } p_{i}^{2}
\ln  V_{i}^{\prime }(0)}  
\]
\[
\times \varepsilon^{(1)}_{\lambda} \varepsilon^{(2)}_{\mu} 
\varepsilon^{(3)}_{\nu}
\left\{ \left( 2 \alpha ^{\prime }\right)^{\frac{3}{2}} \left[
\left( \frac{p_{1}^{\lambda }}{2} \frac{V_{1}^{\prime \prime }(0)}
{V_{1}^{\prime }(0)^{2}}+\frac{p_{2}^{\lambda }}{(z_{1}-z_{2})}+\frac{%
p_{3}^{\lambda }}{(z_{1}-z_{3})}\right) \right. \right.
\]
\[
\times \left( \frac{p_{2}^{\mu}}{2}%
\frac{V_{2}^{\prime \prime }(0)}{V_{2}^{\prime}(0)^{2}}+\frac{p_{1}^{\mu }}{%
(z_{2}-z_{1})}+\frac{p_{3}^{\mu }}{(z_{2}-z_{3})} \right)
\]
\[
\times \left. \left( \frac{p_{3}^{\nu }}{2}\frac{V_{3}^{\prime \prime }(0)}{%
V_{3}^{\prime }(0)^{2}}+\frac{p_{1}^{\nu }}{(z_{3}-z_{1})}+\frac{p_{2}^{\nu }%
}{(z_{3}-z_{2})}\right) \right] 
\]
\[
+\left( 2 \alpha^{\prime} \right)^{\frac{1}{2}}  
\left[ \frac{\eta ^{\lambda \mu }}{(z_{1}-z_{2})^{2}}\left( \frac{%
p_{3}^{\nu }}{2}\frac{V_{3}^{\prime \prime }(0)}{V_{3}^{\prime }(0)^{2}}+%
\frac{p_{1}^{\nu }}{(z_{3}-z_{1})}+\frac{p_{2}^{\nu }}{(z_{3}-z_{2})}\right)
\right. 
\]
\[
 +\frac{\eta ^{\lambda \nu }}{(z_{1}-z_{3})^{2}}\left( \frac{%
p_{2}^{\mu }}{2}\frac{V_{2}^{\prime \prime }(0)}{V_{2}^{\prime }(0)^{2}}%
+
\frac{p_{1}^{\mu }}{(z_{2}-z_{1})}+\frac{p_{3}^{\mu }}{(z_{2}-z_{3})%
}\right) 
\]
\begin{equation}
\left. \left. +\frac{\eta ^{\mu \nu }}{(z_{2}-z_{3})^{2}} \left( 
\frac{p_{1}^{\lambda }}{2}\frac{V_{1}^{\prime \prime }(0)}{V_{1}^{\prime
}(0)^{2}}+\frac{p_{2}^{\lambda }}{(z_{1}-z_{2})}+\frac{p_{3}^{\lambda }}{%
(z_{1}-z_{3})}\right) \right] \right\} .  \nonumber
\label{pippo}
\end{equation}

In the low-energy limit the only contribution that survives is: 
\[
A_{0}^{3}(\alpha ^{\prime } \rightarrow 0)=4g_d\,\,\varepsilon^{(1)}_{\lambda
}\varepsilon^{(2)}_{\mu }\varepsilon^{(3)}_{\nu }\left\{ \frac{\eta ^{\lambda
\mu }}{(z_{1}-z_{2})^{2}}\left( \frac{p_{3}^{\nu }}{2}\frac{V_{3}^{\prime
\prime }(0)}{V_{3}^{\prime }(0)^{2}}+\frac{p_{1}^{\nu }}{(z_{3}-z_{1})}+%
\frac{p_{2}^{\nu }}{(z_{3}-z_{2})}\right) \right.   \nonumber
\]
\[  
+\frac{\eta ^{\lambda \nu }}{(z_{1}-z_{3})^{2}} \left( \frac{p_{2}^{\mu }}{2}%
\frac{V_{2}^{\prime \prime }(0)}{V_{2}^{\prime }(0)^{2}}+\frac{%
p_{1}^{\mu }}{(z_{2}-z_{1})}+\frac{p_{3}^{\mu }}{(z_{2}-z_{3})}\right)  
\]
\[
+\frac{\eta ^{\mu \nu }}{(z_{2}-z_{1})^{2}}\left. \left( \frac{%
p_{1}^{\lambda }}{2}\frac{V_{1}^{\prime \prime }(0)}{V_{1}^{\prime }(0)^{2}}+%
\frac{p_{2}^{\lambda }}{(z_{1}-z_{2})}+\frac{p_{3}^{\lambda }}{(z_{1}-z_{3})}%
\right) \right\} 
\]
\begin{equation}
\times (z_{1}-z_{2})(z_{1}-z_{3})(z_{2}-z_{3}).  \label{ca}
\end{equation}

This amplitude results to be projective invariant if it is satisfied the
following constraint for $V_{i}^{\prime \prime }(0)/(V_{i}^{\prime
}(0))^{2} $
\begin{equation}
\frac{V_{i}^{\prime \prime }(0)}{(V_{i}^{\prime }(0))^{2}}=2\frac{\left(
z_{i}-z_{i+1}\right) -\ell \left( z_{i+1}-z_{i-1}\right) }{\left(
z_{i}-z_{i+1}\right) \left( z_{i}-z_{i-1}\right) }  \label{13}
\end{equation}
where $\ell $ is a free parameter. Then there exist, as
solutions, a set of
functions of the Koba-Nielsen variables parametrized by $\ell$. In order
to choose a
value for that parameter, we impose our low-energy amplitude 
to reproduce a gauge independent result such as the one relative to the 
three-photon scattering amplitude which is identically zero. For extracting 
this amplitude 
we sum the expression (\ref{ca}) over all the anticyclic 
permutations of the indices $(1,2,3)$ getting: 
\begin{equation}
A_{0}^{3}(photons)=4g_d\,\,\varepsilon^{(1)}_{\lambda }\varepsilon^{(2)}_{\mu
}\varepsilon^{(3)}_{\nu }(2\ell +1)\left\{ \eta ^{\lambda \mu }(p_{1}^{\nu
}+p_{2}^{\nu })+\eta ^{\mu \nu }(p_{3}^{\lambda }+p_{2}^{\lambda })+\eta
^{\lambda \nu }(p_{1}^{\mu }+p_{3}^{\mu })\right\} =0  \label{14}
\end{equation}
which is satisfied only for 
\begin{equation}
\ell =-\frac{1}{2} .  \label{15}
\end{equation}
With this value of the parameter $\ell $ the eq. (\ref{13}) becomes 
\begin{equation}
\frac{V_{i}^{\prime \prime }(0)}{(V_{i}^{\prime }(0))^{2}}=\frac{1}{\left(
z_{i}-z_{i+1}\right) }+\frac{1}{\left( z_{i}-z_{i-1}\right) } . \label{16}
\end{equation}
It is easy to check that the Lovelace function does not satisfy the
equation (\ref{16}). By substituting (\ref{16}) into (\ref{ca}) we get the
following expression of the
low-energy limit of the amplitude in consideration:
\begin{equation}
A_{0}^{3}(\alpha^{\prime }\rightarrow 0)=2g_{d}\,\,\varepsilon^{(1)}_{\lambda
}\varepsilon^{(2)}_{\mu }\varepsilon^{(3)}_{\nu }\left\{ \eta ^{\lambda \mu
}\left( p_{1}^{\nu }-p_{2}^{\nu }\right) +\eta ^{\lambda \nu }\left(
p_{3}^{\mu }-p_{1}^{\mu }\right) +\eta ^{\mu \nu }\left( p_{2}^{\lambda
}-p_{3}^{\lambda }\right) \right\} . \label{17}
\end{equation}
The choice (\ref{16}) makes the entire amplitude (%
\ref{pippo}) to become:
\begin{eqnarray}
A_{0}^{3} &=&g_d(z_{1}-z_{2})^{ 2\alpha ^{\prime } p_{1\cdot
}p_{2}}(z_{1}-z_{3})^{ 2\alpha ^{\prime } p_{1\cdot
}p_{3}}(z_{2}-z_{3})^{ 2\alpha ^{\prime } p_{2\cdot
}p_{3}}  \label{18} \\
&& \times \prod\limits_{i=1}^{3} V_{i}^{\prime }(0)^{
\alpha^{\prime} p_{i}^{2}} \varepsilon^{(1)}_{\lambda }
\varepsilon^{(2)}_{\mu }\varepsilon^{(3)}_{\nu } \left\{ \left( 2\alpha ^{\prime }\right) \left[ \frac{1}{2}\left(
p_{2}^{\lambda }-p_{3}^{\lambda }\right) \left( p_{3}^{\mu }-p_{1}^{\mu
}\right) \left( p_{1}^{\nu }-p_{2}^{\nu }\right) \right] +\right.  \nonumber
\\
&&+2\left. \left[ \eta ^{\lambda \mu }\left( p_{1}^{\nu }-p_{2}^{\nu
}\right) +\eta ^{\lambda \nu }\left( p_{3}^{\mu }-p_{1}^{\mu }\right) +\eta
^{\mu \nu }\left( p_{2}^{\lambda }-p_{3}^{\lambda }\right) \right] \right\}. 
\nonumber
\end{eqnarray}
If this latter amplitude has to be projective invariant, then 
$V_{i}(z)$ must satisfy the following equation: 
\begin{equation}
(z_{1}-z_{2})^{ 2\alpha ^{\prime }p_{1\cdot }p_{2}}(z_{1}-z_{3})^{
 2\alpha ^{\prime } p_{1\cdot }p_{3}}(z_{2}-z_{3})^{ 2\alpha
^{\prime } p_{2\cdot }p_{3}} \prod\limits_{i=1}^{3} 
\left[V_{i}^{\prime
}(0)\right]^{ \alpha ^{\prime } p_{i}^{2}} = const.  \label{19}
\end{equation}
From (\ref{16}) we know that $V_{i}^{\prime }(0)$
must be a function of all the punctures, then we can conjecture it to be
written as follows: 
\begin{equation}
V_{i}^{\prime }(0)=\left( z_{i}-z_{i+1}\right) ^{a}\left(
z_{i}-z_{i-1}\right) ^{b}\left( z_{i+1}-z_{i-1}\right) ^{c}  \label{19 bis}
\end{equation}
where $\ a$, $b$, $c$, are unknown exponents to be determined by resolving
the eq. (\ref{19}). Indeed, by inserting (\ref{19 bis}) into (\ref{19}) we 
get the following system equations for $a$, $b$, $c$:
\begin{equation}
\left\{ 
\begin{array}{l}
p_{1}^{2}a+p_{2}^{2}b+p_{3}^{2}c=-2p_{1}\cdot p_{2} \\ 
p_{3}^{2}a+p_{1}^{2}b+p_{2}^{2}c=-2p_{1}\cdot p_{3} \\ 
p_{2}^{2}a+p_{3}^{2}b+p_{1}^{2}c=-2p_{3}\cdot p_{2} .
\end{array}
\right.  \label{19 ter}
\end{equation}
This system admits a unique solution that, using the momentum conservation, 
is:
\begin{equation}
\left\{ 
\begin{array}{l}
a=1 \\ 
b=1 \\ 
c=-1 .
\end{array}
\right.  \label{19 quarter}
\end{equation}
Plugging it into (\ref{19 bis}) yields: 
\begin{equation}
V_{i}^{\prime }(0)=\frac{(z_{i}-z_{i+1})(z_{i}-z_{i-1})}{(z_{i+1}-z_{i-1})}.
\smallskip  \label{10}
\end{equation}
It implies that all the tree Green functions involved 
in the three-string case and defined by eq. (\ref{8bis}) vanish. The 
eq. (\ref{10}) corresponds
to the first derivative of the Lovelace function $V_{i}(z)$
evaluated at $z=0$ \cite{L}.
  
By requiring, then, the projective invariance
of the low-energy limit of the amplitude (\ref{ca}) and it
to reproduce the three-photon scattering amplitude, we could fix the
ratio $V_{i}^{\prime \prime }(0)/(V_{i}^{\prime }(0))^{2}$; at the
same time, by requiring the projective invariance of the scattering
amplitude (\ref{ca}) we get a specific prescription for the
function $V_{i}^{\prime}(0)$. If the local holomorphic
coordinate system $V_{i}(z)$ is chosen to be a projective transformation 
satisfying
by construction the condition $V_{i}(0)=z_{i}$, we
completely fix the function $V_{i}(z)$ by fixing $V_{i}^{\prime
}(0) $ and $V_{i}^{\prime \prime }(0)/(V_{i}^{\prime }(0))^{2}$.

Introducing the choices (\ref{10}) and (\ref{16}) in (\ref{18}) 
gives
the following projective invariant three-open string off-shell
amplitude: 
\begin{eqnarray}
A_{0}^{3} &=&g_d \varepsilon^{(1)}_{\lambda }\varepsilon^{(2)}_{\mu }
\varepsilon^{(3)}_{\nu }\left\{ \left( 2 \alpha^{\prime } \right) \left[ \frac{1}{%
2}\left( p_{3}^{\lambda }-p_{2}^{\lambda }\right) \left( p_{1}^{\mu
}-p_{3}^{\mu }\right) \left( p_{2}^{\nu }-p_{1}^{\nu }\right) \right]
\right. +  \label{20} \\
&&+2\left. \left[ \eta ^{\lambda \mu }\left( p_{2}^{\nu }-p_{1}^{\nu
}\right) +\eta ^{\lambda \nu }\left( p_{1}^{\mu }-p_{3}^{\mu }\right) +\eta
^{\mu \nu }\left( p_{3}^{\lambda }-p_{2}^{\lambda }\right) \right] \right\}. 
\nonumber
\end{eqnarray}

Let us now come back to the analysis of the field theory limit of the open
string amplitude. We are interested in the evaluation of the three-gluon
scattering amplitude. As it is well-known, at tree level we need to sum the
expression in (\ref{17}) over all the anticyclic permutations of the indices $%
(1,2,3)$ after having multiplied it by the Chan-Paton factor 
\begin{equation}
\text{Tr }\left( \lambda _{a_{1}}\text{ }\lambda _{a_{2}}\text{ }\lambda
_{a_{3}}\right)  \label{21}.
\end{equation}
In this object $\lambda$'s are the generators of the $SU(N)$ gauge group
in the fundamental representation. Through the normalization conditions 
\begin{equation}
\text{Tr }\left( \lambda _{a}\text{ }\lambda _{b}\right) =\frac{1}{2}\delta
_{ab}  \label{22}
\end{equation}
the Chan-Paton factor can be rewritten as: 
\begin{equation}
\text{Tr }\left( \lambda _{a_{1}}\text{ }\lambda _{a_{2}}\text{ }\lambda
_{a_{3}}\right) =\frac{1}{4}\left(
f^{a_{1}a_{2}a_{3}}+d^{a_{1}a_{2}a_{3}}\right),  \label{23}
\end{equation}
where $f$ is an antisymmetric tensor with respect the internal indices $a_{i}$ and $d$
a symmetric one. An explicit evaluation of the three-gluon amplitude yields: 
\begin{equation}
A_{0}^{3}(\text{gluons})=g_d\,\,\varepsilon^{(1)}_{\lambda }
\varepsilon^{(2)}_{\mu }\varepsilon^{(3)}_{\nu }f^{abc}\left\{ \eta ^{\lambda \mu }\left(
p_{1}^{\nu }-p_{2}^{\nu }\right) +\eta ^{\lambda \nu }\left( p_{3}^{\mu
}-p_{1}^{\mu }\right) +\eta ^{\mu \nu }\left( p_{2}^{\lambda
}-p_{3}^{\lambda }\right) \right\}.  \label{24}
\end{equation}
This expression coincides with the one of the three-gluon scattering
amplitude obtained through a covariant quantization of a non abelian gauge
theory with gauge group $SU(N)$ or by using the background field method.

In computing a string amplitude in our formalism, the choice of 
the function $V_{i}(z)$ seems to be strictly connected with the gauge
choice in the field theory limit of those amplitudes. Since the request of
consistency, like that we have here made, together with the conjecture 
(\ref{19 bis}), completely fixes $V_{i}(z),$ it 
turns out that the only way in which the open string can reproduce the 
field theory gauge-dependent result of the
three-gluon scattering amplitude at tree level is in the usual covariant gauge
 \cite{DMLRM}.

Let us analyze the closed string case, by extending to it the
result in (\ref{17}). 
We find that the scattering amplitude involving three external off-shell
massless states admits the following low-energy limit:

\begin{eqnarray}
A_{0}^{3c}(\alpha ^{\prime } \rightarrow 0) = & k \,\,\varepsilon
^{(1)}_{\lambda \rho }\varepsilon^{(2)}_{\mu \sigma }
\varepsilon^{(3)}_{\nu \tau
}\left\{ \left[ \eta ^{\lambda \mu }\left( p_{1}^{\nu }-p_{2}^{\nu }\right)
+\eta ^{\lambda \nu }\left( p_{3}^{\mu }-p_{1}^{\mu }\right) +\eta ^{\mu \nu
}\left( p_{2}^{\lambda }-p_{3}^{\lambda }\right) \right] \right. \nonumber
\\
 {} & \times \left. \left[ \eta ^{\rho \sigma }\left( p_{1}^{\tau }-p_{2}^{\tau
}\right) +\eta ^{\rho \tau }\left( p_{3}^{\sigma }-p_{1}^{\sigma }\right)
+\eta ^{\sigma \tau }(p_{2}^{\rho }-p_{3}^{\rho })\right] \right\}.
\end{eqnarray}
After symmetrizing this expression to eliminate the contribution of the
antisymmetric tensor, we get the scattering amplitude of three gravitons mixed
with dilatons: 
\[
A_{0}^{3c} (_{\text{dilatons}}^{\text{gravitons}})= k \,\,\varepsilon
_{\lambda \rho }^{(1)}\varepsilon _{\mu \sigma}^{(2)}
\varepsilon_{\nu \tau}^{(3)}\left\{ I^{\lambda \rho \mu \sigma }\left( p_{2}^{\nu }-p_{1}^{\nu
}\right) \left( p_{2}^{\tau }-p_{1}^{\tau }\right) +I^{\lambda \rho \nu \tau
}\left( p_{1}^{\mu }-p_{3}^{\mu }\right) \left( p_{1}^{\sigma
}-p_{3}^{\sigma }\right) \right.
\]
\[ 
+I^{\mu \sigma \nu \tau }\left( p_{3}^{\lambda }-p_{2}^{\lambda }\right)
\left( p_{3}^{\rho }-p_{2}^{\rho }\right) +\frac{1}{2}\left[ I^{\lambda \rho
\mu \tau }\left( p_{2}^{\nu }-p_{1}^{\nu }\right) \left( p_{1}^{\sigma
}-p_{3}^{\sigma }\right) \right. \]
\[
+I^{\lambda \rho \sigma \tau }\left( p_{2}^{\nu }-p_{1}^{\nu }\right)
\left( p_{1}^{\mu }-p_{3}^{\mu }\right) +I^{\lambda \rho \mu \nu }\left(
p_{1}^{\sigma }-p_{3}^{\sigma }\right) \left( p_{2}^{\tau }-p_{1}^{\tau
}\right) +I^{\lambda \rho \sigma \nu }\left( p_{1}^{\mu }-p_{3}^{\mu
}\right) \left( p_{2}^{\tau }-p_{1}^{\tau }\right) 
\]
\[
+I^{\lambda \tau \mu \sigma }\left( p_{2}^{\nu }-p_{1}^{\nu }\right)
\left( p_{3}^{\rho }-p_{2}^{\rho }\right) +I^{\rho \tau \mu \sigma }\left(
p_{2}^{\nu }-p_{1}^{\nu }\right) \left( p_{3}^{\lambda }-p_{2}^{\lambda
}\right) +I^{\lambda \nu \mu \sigma }\left( p_{3}^{\rho }-p_{2}^{\rho
}\right) \left( p_{2}^{\tau }-p_{1}^{\tau }\right)  
\]
\[
+I^{\rho \nu \mu \sigma }\left( p_{3}^{\lambda }-p_{2}^{\lambda }\right)
\left( p_{2}^{\tau }-p_{1}^{\tau }\right)  
+I^{\lambda \sigma \nu \tau }\left( p_{1}^{\mu }-p_{3}^{\mu
}\right) \left( p_{3}^{\rho }-p_{2}^{\rho }\right) +
I^{\lambda \mu \nu \tau }\left(p_{1}^{\sigma }-p_{3}^{\sigma}
\right)
\left( p_{3}^{\rho }-p_{2}^{\rho }\right)  
\]
\begin{equation}
\left. \left. +I^{\rho \sigma \nu \tau }\left(
p_{1}^{\mu }-p_{3}^{\mu }\right) \left( p_{3}^{\lambda }-p_{2}^{\lambda
}\right) +I^{\rho \mu \nu \tau }\left( p_{1}^{\sigma }-p_{3}^{\sigma
}\right) \left( p_{3}^{\lambda }-p_{2}^{\lambda } \right) \right] \right\}
                \label{3grsc} 
\end{equation}
with
\begin{equation}
I_{\alpha \beta \gamma \delta}= \frac{1}{2} \left( \eta_{\alpha \gamma}
\eta_{\beta \delta} + \eta_{\alpha \delta} \eta_{\beta \gamma} \right).
\label{ID}
\end{equation}

Although the De Donder gauge is commonly believed to be the preferred gauge 
choice in quantum gravity \cite{HV}, there are some arguments \cite{BDDPR} 
which suggest that it is not the one chosen by the bosonic
closed string. 

We have investigated this point by computing 
the expansion of the 
Einstein-Hilbert Lagrangian coupled to a dilaton field with the De Donder 
gauge fixing term, up to the third order in the fields. On the complete 
Lagrangian we have performed the following transformations \cite{BDS}:  

\beqa
h_{\mu \nu} = \tilde{h}_{\mu \nu} +  \frac{\eta_{\mu \nu}}{\sqrt{D-2}} 
\tilde{\phi}\\ 
\phi = \sqrt{ \frac{D-2}{2}} \tilde{\phi} + \frac{1}{\sqrt{2}} 
\tilde{h}^{\mu}_{\mu} .
\eeqa

These have been determined by requiring that the propagator for
$h_{\mu \nu}$ 
generated by the transformed Lagrangian were proportional to the 
unit tensor given in (\ref{ID}), that has a trivial Lorentz structure
compatibly with the form of the string propagator which does not contain
any Lorentz indices \cite{BDS}.
Then the computation of the three-point function from the Lagrangian 
so obtained has not reproduced the result shown in (\ref{3grsc}).
 
This means that,
within this comprehension of decoupling gravitons and dilatons, and in 
agreement with the arguments of ref. \cite{BDDPR},   
the closed string {\em does not choose, in the low-energy limit, the
De Donder gauge}, which, naively, might be considered as the most natural
counterpart of the usual covariant gauge in the open string case.

\vspace{2cm}

The authors thank L. Cappiello and R. Pettorino for helpful
discussions and for a critical reading of a preliminary version of the
paper. 

A.L. thanks NORDITA for the kind hospitality; R.M. and F.P. respectively 
thank the Dipartimento di Scienze Fisiche dell'Universit\`{a} di Napoli 
and NORDITA for the hospitality kindly provided to them
in different stages of this work.


\begin{thebibliography}{99}

\bibitem{BK}  Z. Bern and D.A. Kosower, {\em Phys. Rev.} {\bf D38}
(1988) 1888; {\em Nucl. Phys.} {\bf B321} (1989) 605; {\em Nucl. Phys.} 
{\bf B379} (1992) 451.

\bibitem{DMLRM}  P. Di Vecchia, R. Marotta, A. Lerda, R. Russo and L. Magnea, 
{\em Nucl. Phys.} {\bf B469} (1996) 235.

\bibitem{BDS}  Z. Bern, D.C. Dunbar and T. Shimada, {\em Phys. Lett.} 
{\bf B312} (1983) 277.

\bibitem{P} J. Polchinski, {\em Phys. Rev. Lett.} {\bf 75} (1995) 4727, 
hep-th/9510017; 

 S. Ramgoolam and L. Thorlacius, {\em Nucl. Phys.} {\bf B483}
(1997) 248, hep-th/9607113; 

 M. Bill\'o, P. Di Vecchia, M. Frau, A. Lerda, I. Pesando, R. Russo and 
S. Sciuto, {\em Nucl. Phys.} {\bf B526} (1998) 199, hep-th/9802088.

\bibitem{CF} E. F. Corrigan and D. B. Fairlie, {\em Nucl. Phys.} {\bf B91}
(1975) 527.

\bibitem{CMNP} A. Cohen, G. Moore, P. Nelson and J. Polchinski, 
{\em Nucl. Phys.}{\bf B281} (1987) 127.

\bibitem{LS} S. Samuel, {\em Nucl. Phys.} {\bf B308} (1988) 285; 
O. Lechtenfeld and S. Samuel,
{\em Nucl. Phys.} {\bf B308} (1988) 361.

\bibitem{BKR} Z. Bern, D.A. Kosower and K. Roland,
{\em Nucl. Phys.} {\bf B344} (1990) 309.

\bibitem{MP} R.C. Myers and V. Periwal, Talk given at the International
Workshop on String Theory, Quantum Gravity and the Unification of Fundamental
Interactions, Rome, 1992, hep-th/9211078.

\bibitem{BZ} A. Belopolsky and B. Zwiebach, {\em Nucl. Phys.} {\bf B442}
(1995) 494, hep-th/9409015.

\bibitem{DPFLHS}  P. Di Vecchia, F. Pezzella, M. Frau, A. Lerda, K.
Hornfeck and S. Sciuto, {\em Nucl. Phys.} {\bf B322} (1989) 317.

\bibitem{CMPP1}  L. Cappiello, R. Marotta, R. Pettorino and F. Pezzella, 
{\em Mod. Phys. Lett.} {\bf A13} (1998) 2433, hep-th/9804032.

\bibitem{CMPP2} L. Cappiello, R. Marotta, R. Pettorino and F. Pezzella,
{\em Mod. Phys. Lett.} {\bf A13} (1998) 2875, hep-th/9808164. 


\bibitem{KV} T. Kubota and G. Veneziano, {\em Phys. Lett.} {\bf B207} (1988)
419.

\bibitem{L} C. Lovelace, {\em Phys. Lett.} {\bf B32} (1970) 490.

\bibitem{RS}  K. Roland and H. Sato, {\em Nucl. Phys.} {\bf B480} (1996) 99, 
hep-th/9604152; {\em Nucl. Phys.} {\bf B515} (1998) 488, hep-th/9709019.

\bibitem{BD} Z. Bern and D.C. Dunbar, {\em Nucl. Phys.} {\bf B379} (1992) 451.

\bibitem{KLT} H. Kawai, D.C. Lewellen and S.-H.H. Tye, {\em Nucl. Phys.}
{\bf B269} (1986) 1. 

\bibitem{HV} G. 't Hooft and M. Veltman, {\em Ann. Inst. Henri Poincar\`{e}},
{\bf B266}, 799 (1986).

\bibitem{BDDPR} Z. Bern, L. Dixon, D.C. Dunbar, M. Perelstein and J.S. 
Rozowsky, {\em Perturbative Relationships Between QCD and Gravity and Some
Implications}, hep-th/9809163.  

\end{thebibliography}
\end{document}